# Unconventional ferroelectric domain switching dynamics in $CuInP_2S_6$ from first principles

Ri He[1], Hua Wang[2], Fucai Liu[3,4], Shi Liu[5,6], Houfang Liu[7], Zhicheng Zhong[1,8]

[1]Key Laboratory of Magnetic Materials Devices & Zhejiang Province Key Laboratory of Magnetic Materials and Application Technology, Ningbo Institute of Materials Technology and Engineering, Chinese Academy of Sciences, Ningbo 315201, China
[2]School of Micro-Nano Electronics, Hangzhou Global Scientific and Technological Innovation Center, Zhejiang University, Hangzhou, 310027, China
[3]School of Optoelectronic Science and Engineering, University of Electronic Science and Technology of China, Chengdu, 611731, China
[4]Yangtze Delta Region Institute (Huzhou), University of Electronic Science and Technology of China, Huzhou, 313001, China
[5]Key Laboratory for Quantum Materials of Zhejiang Province, Department of Physics, School of Science, Westlake University, Hangzhou, 310024, China
[6]Institute of Natural Sciences, Westlake Institute for Advanced Study, Hangzhou 310024, China
[7]Institute of Microelectronics and Beijing National Research Center for Information Science and Technology (BNRist), Tsinghua University, Beijing 100084, China
[8]China Center of Materials Science and Optoelectronics Engineering, University of Chinese Academy of Sciences, Beijing 100049, China

**Abstract**

The switching dynamics of ferroelectric materials is a crucial intrinsic property which directly affects the operation and performance of ferroelectric devices. In conventional ferroelectric materials, the typical ferroelectric switching mechanism is governed by a universal process of domain wall motion. However, recent experiments indicate that Van der Waals ferroelectric $CuInP_2S_6$ possesses anomalous polarization switching dynamics under an electric field. It is important to understand the switching dynamics, but it remains theoretically unexplored in $CuInP_2S_6$ due to the lack of description of its order-disorder phase transition characteristics by density functional theory. Here, we employ a machine-learning potential trained from the first principles density functional theory dataset to conduct the large-scale atomistic simulations of temperature-driven order-disorder ferroelectric phase transition in $CuInP_2S_6$. Most importantly, it is found that the electric field-driven polarization switching in $CuInP_2S_6$ is mediated by single Cu dipole flip, rather than conventional domain wall motion mechanism. This intrinsic unconventional switching behavior can be attributed to the competition between the energy barrier of domain wall motion and single dipole flip.

# Introduction

Van der Waals (vdW) layered ferroelectric materials have recently become an emerging research branch in ferroelectric physics, providing an opportunity to explore ferroelectricity at the two-dimensional limit[1-3]. Several vdW ferroelectrics with intrinsic either in-plane or out-of-plane polarization have been discovered to date, including $CuInP_2S_6$ (CIPS)[4,5], $In_2Se_3$[6], $MoTe_2$[7], and single-element Bi[8] etc. CIPS is one of the most representative materials because of its rich properties, such as robust room temperature ferroelectricity[4], finite ionic conductivity[9], photovoltaic effect[10], electrocaloric effect[11], and giant negative piezoelectricity[12]. Besides, CIPS is the only order-disorder ferroelectrics in vdW materials. At low temperatures, CIPS has a ferroelectric monoclinic structure with *Cc* space group. The out-of-plane spontaneous polarization originates from Cu cations favoring lower coordination either at upper or bottom instead of sitting at the center of the sulfur octahedron as illustrated in Fig. 1. Above the Curie temperature of 315 K, thermal fluctuations drive Cu dipoles to randomly flip between upper or bottom sites, leading to a macroscopically nonpolar paraelectric phase with a spatial disordered Cu sublattice[13,14].

Such order-disorder ferroelectric phase transition mediated by Cu cations hopping motions may result in an anomalous polarization switching dynamics under an electric field[9]. Conventional ferroelectrics exhibit a good linear relationship between the switching time and reciprocal of the electric field, according to the Merz's law[15]. However, Zhou *et al.* recently discovered that the CIPS showed a strong deviation from this linear relationship, which is not in line with any previously proposed models[9]. Although they attributed this anomalous behavior to the interaction between polarization and ionic defect dipoles, the anomalous switching is largely governed by an unusual mechanism of intrinsic Cu hopping motions, even in the absence of defects[16].

Indeed, switching dynamics is one of the extremely important features for ferroelectric materials as they directly affects the performance of ferroelectric devices. Polarization switching dynamics in conventional ferroelectrics, such as $BaTiO_3$,

PbZr$_x$Ti$_{(1-x)}$O$_3$, have been extensively studied experimentally and theoretically[15,17-20]. It is suggested that the typical ferroelectric switching is achieved by a simple, universal mechanism of intrinsic domain wall motion[19,20]. However, the anomalous switching in CIPS indicates an unusual mechanism of domain switching that remains unknown. Unfortunately, due to the time/spatial resolution of the piezoresponse force microscopy, it is difficult to figure out domain switching process experimentally. Theoretically, density functional theory (DFT) is unable to describe its temperature-driven order-disorder ferroelectric phase transition due to the simulation's spatial and time limitation. The classical force field is not available for such a multi-component and structurally complicated material like CIPS, making the direct classical molecular dynamics (MD) simulation impossible.

In this article, we report a machine-learning based deep potential (DP) model, trained using DFT calculations, to systematically and directly investigate the order-disorder phase transition and domain switching process in vdW ferroelectrics CIPS from first principles. The well-trained DP model accurately predicted the static and dynamic properties of CIPS, such as lattice constants, phonon spectrum, domain wall energies, and ferroelectric switching barriers, with DFT-level accuracy. Using the DP model, we perform deep potential molecular dynamic (DPMD) simulations to capture the temperature-driven order-disorder ferroelectric-paraelectric phase transition and the Cu cations hopping motions in CIPS. The calculated $T_C$ = 340 K is in good agreement with the experimental value of 315 K. In addition, the simulation results indicate the out-of-plane lattice constant of bulk CIPS exhibits a large negative thermal expansion coefficient with $-9.85 \times 10^{-6}$ K$^{-1}$ below 180 K, which originates from the nuclear quantum effect. Most importantly, we demonstrate that the polarization switching in CIPS is mediated by single Cu dipole flip, rather than conventional domain wall motion. This unconventional switching dynamics may shed light on the nature of ferroelectric switching in order-disorder ferroelectric materials.

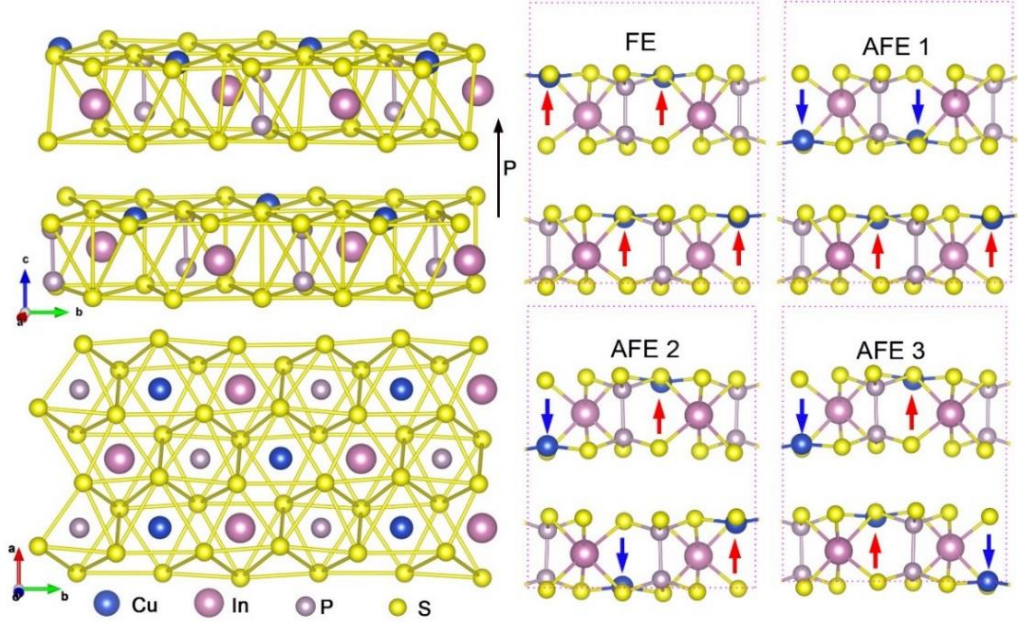

Figure 1: (a) Top and side views of monoclinic structure of ferroelectric CIPS. (b) Ferroelectric (FE) and various possible antiferroelectric (AFE) ordering, the arrows represent the direction of local polarization.

# Method

### A. Machine-learning interatomic potential

To quantitatively analyze the polarization switching process in multi-domain CIPS with a sufficiently large time and spatial scale, density functional theory (DFT)-accuracy atomistic dynamics simulations at nanosecond and million atom scale are required. Machine-learning interatomic potential models is emerging methods that describe atomic interactions, bridging the gap between accurate first principles DFT calculations and efficient large-scale atomistic dynamics simulations[21,22]. Deep potential (DP) proposed in recent years is one of the most widely used deep neural network type potential[23,24]. Although still in its infancy, DP method has been successfully applied to investigate ferroelectric-related properties in several vdW ferroelectrics, such as monolayer $In_2Se_3$[25] and sliding/Moiré ferroelectric bilayer BN[26]. Herein, using the DP method we train a deep neural network potential of layered CIPS based on first-principles DFT calculation data according to the following steps.

Selecting appropriate atomic configurations for the training dataset is crucial for the accuracy of DP model. Our focus is on the studying the dynamics response of the multi-domain structure at finite temperature and under an external field, which requires inclusion of all characteristic structures, including intermediate states of polarization switching and lattice distortion. Manual construction of these configurations is a tedious task, particularly for the states far from the equilibrium configuration. Therefore, we adopt a concurrent learning procedure that automatically generates training data, covering the entire relevant configuration space[27]. The workflow of each iteration in concurrent learning procedure includes three main steps: (1) training the DP model, (2) exploring configurations by running *NPT* deep potential molecular dynamics simulations at different temperatures, and (3) labeling configurations based on specific criteria, adding them to the training dataset, and then repeating the step 1 again. We start with 1×1×1 and 1×2×2 supercells of ferroelectric and various antiferroelectric (AFE1, AFE2, and AFE3) structures as shown in Fig. 1, as well as intermediate polarization states optimized by DFT calculations. To simulate the CIPS thin-films with different thickness, the mono-, bi-, triple-, and quadruple-layer configurations were included in the training dataset. 11260 training configurations were generated after iterating this procedure 23 times. The DP model is trained using these configurations and corresponding PBE-based DFT energies, with fitting networks of size (240, 240, 240)[24].

## B. Density functional theory calculations

The initial training dataset is obtained by performing a 20-step *ab-initio* MD simulation for randomly perturbed structures at 50 K. Once all configurations are collected in final training dataset, the self-consistent DFT would be performed. The *ab-initio* MD and self-consistent DFT calculations were performed using a plane-wave basis set with a cutoff energy of 550 eV as implemented in the Vienna Ab initio Simulation Package [28,29]. To obtain an accurate deep potential model for Van der Waals layered CIPS, we employ the Van der Waals correction with the optB86b exchange-correlation function in DFT calculations[30]. The Brillouin zone was sampled with a 10 × 8 × 2 *k*-point grid for 40-atom monoclinic unit cell. In the CIPS flake calculations, a vacuum region of 40 Å is included to prevent the coupling between periodic images. To reproduce the domain wall motion from initial equilibrium state to another, the climbing image nudged elastic band (NEB) method was used to find the minimum energy path for the domain wall motion[31].

## C. Molecular dynamics simulations

Once the well-trained DP model is obtained, the deep potential molecular dynamics (DPMD) simulations can be conducted to investigate the temperature and electric field-driven polarization switching process in CIPS. The MD simulations were carried out using LAMMPS code with periodic boundary conditions[32]. In temperature-driven order-disorder phase transition simulation, we used a sufficiently large 38400-atom supercell consisting of 15 × 8 × 8 unit cells to perform *NPT* MD simulations. Temperature and pressure are controlled by a Nose-Hoover thermostat and Parrinello-Rahman barostat respectively[32].The time step is set to 0.001 ps. To explore the evolution of in-plane domain structure, a 256 000-atom supercell consisting of 80 × 40 × 2 unit cells was used in MD simulations to produce the 48.9 nm × 42.3 nm size of in-plane 2D map. We also used 80 × 40 × 1 and 80 × 40 × 0.5 supercells (including vacuum region of 40 Å) to explore the domain evolution dynamics in CIPS bilayer and monolayer. After reaching the thermodynamic equilibrium, the polarization (*P*) of local formula unit could be calculated by the atomic coordinates displacement ($u_i$) with respect to the referenced non-polar phase multiplied by the Born effective charges tensors ($Z_i^*$)[33]:

$$P = \frac{1}{V}\sum_i Z_i^* u_i$$

where *V* is the volume of formula unit. In the case of CIPS, the presence of covalent P-P and P-S bonds in CIPS precludes assumption of point charges in the calculation. The ethane-like $P_2S_6$ entity can be treated as a rigid ion. We only focus on the out-of-plane polarization, and the out-of-plane component of Born effective charge tensors are

$Z^*_{P2S6} = -2.4$, $Z^*_{Cu} = 0.6$, and $Z^*_{In} = 1.8$, according to previous work[13]. The macro-polarization of supercell at specific temperature can be calculated by averaging all the polarization of local formula unit. Here, we only calculate the out-of-plane polarization, because the in-plane polarization in bulk CIPS is sufficiently small[34].

Table I. The lattice constants and energies (meV) of ferroelectric (FE) and antiferroelectric (AFE) structures in Figure 1 fully relaxed by the deep potential (DP) model and DFT.

| Structures | *a* (Å) | | *b* (Å) | | *c* (Å) | | Energy (meV/f.u.) | |
|---|---|---|---|---|---|---|---|---|
| | DP | DFT | DP | DFT | DP | DFT | DP | DFT |
| Bulk FE | 6.115 | 6.105 | 10.574 | 10.571 | 13.238 | 13.216 | 0 | 0 |
| Bulk AFE 1 | 6.108 | 6.101 | 10.577 | 10.573 | 13.289 | 13.263 | 15.55 | 16.10 |
| Bulk AFE 2 | 6.138 | 6.124 | 10.616 | 10.619 | 13.435 | 13.375 | 30.64 | 22.29 |
| Bulk AFE 3 | 6.136 | 6.122 | 10.612 | 10.624 | 13.344 | 13.297 | 3.33 | 4.67 |
| FE bilayer | 6.108 | 6.101 | 10.574 | 10.564 | - | - | 0 | 0 |
| AFE bilayer | 6.083 | 6.077 | 10.579 | 10.577 | - | - | -24.85 | -21.52 |
| Monolayer | 6.108 | 6.100 | 10.563 | 10.559 | - | - | -15.24 | 0 |

# Results

## A. Deep potential predicted static properties

A high-quality DP model trained from abundance of DFT dataset allows us to directly compare the calculated DFT results. To evaluate the performance of the model, we calculated the energies and atomic forces for all configurations in the final training dataset using both DFT and the DP model. The results, as illustrated in Fig. 2, show excellent agreement between the two methods, with a mean absolute error of energy ($|\Delta E^{DP-DFT}|$) of 1.10 meV/atoms and atomic force ($|\Delta f^{DP-DFT}|$) of 0.70 eV/Å. The DP model accurately predicts structure properties and energies of both the ferroelectric and typical antiferroelectric (AFE) phase structures (Fig. 1), which are in excellent agreement with the results obtained from DFT, as summarized in Table I. Both the DP and DFT results show that the structures are ordered from lowest to highest energy as follows: FE, AFE-3, AFE-1, and AFE-2 (see Fig. 1) for bulk system. However, in the case of the bilayer structure, interlayered AFE ordering is more thermodynamically stable than the ferroelectric ordering, as it helps to screen the large depolarization field.

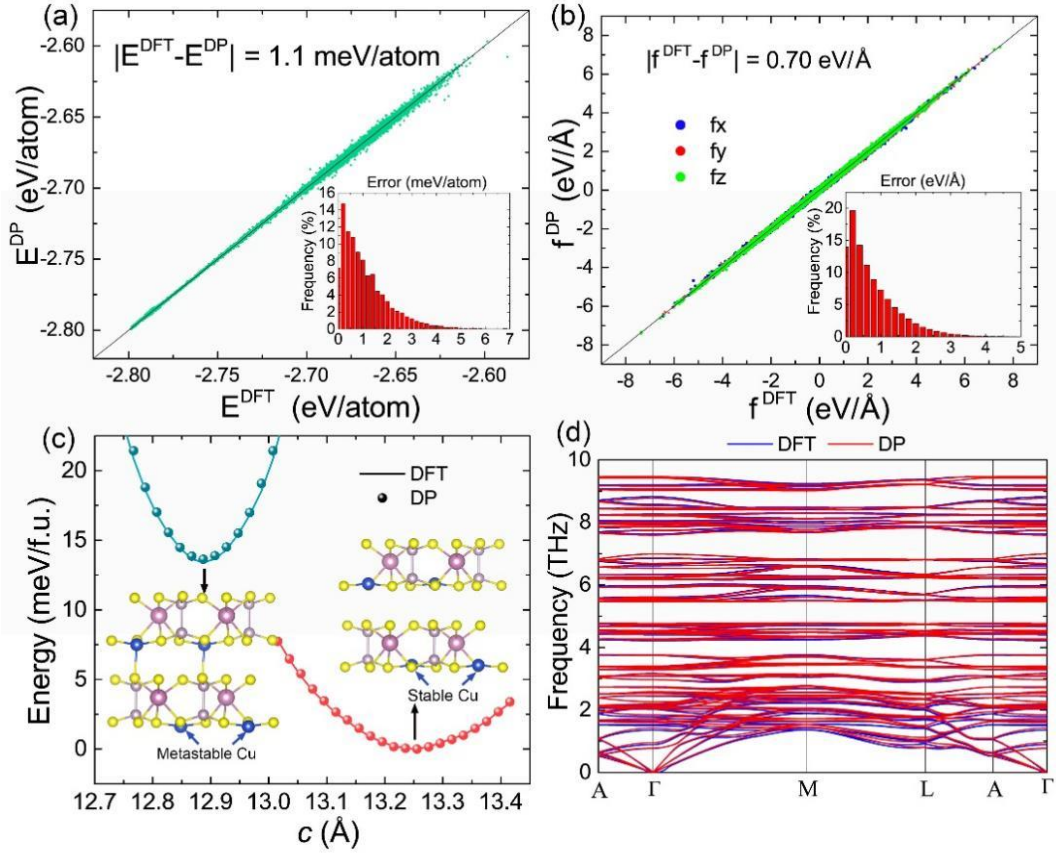

Figure 2: The benchmark test of deep potential model against DFT results. Comparison of energies (a) and atomic force (b) of the DP model against DFT calculations for all configurations in the training dataset. The energies versus out-of-plane lattice for the stable and metastable Cu sites (c) and phonon dispersion relations (d) also calculated by the DP model and DFT.

Recent experiments measuring the electron density in bulk CIPS have suggested the presence of a metastable Cu site within the vdW gap[9]. The DFT calculations confirm this metastable state as illustrated in Fig. 2(c), with an energy difference of 13.41 meV compared to the ground state, which aligns well with the DP prediction of 13.64 meV. The transition of Cu from the ground state to the metastable state also induces a significant contraction of the out-of-plane lattice, resulting in a strain of -2.78%. The DP predicted energy-strain profile for bulk CIPS is plotted in Fig. 2(c), demonstrating that the DP model accurately reproduces the DFT results across a wide range of out-of-plane lattice constants, ranging from 12.7 to 13.4 Å, for both the ground and metastable states, respectively. Additionally, the phonon dispersion of the ferroelectric monoclinic phase is calculated by the DP and DFT as implemented in Phonopy code[35]. As shown in Fig. 2(d), the 40-atom unit cell produces 3 acoustic and 117 optical branches, the three acoustic branches exist only in the low-frequency regime and approach zero frequency near the high-symmetry Gamma point. The DP result exhibits remarkable agreement with DFT curves.

To quantitatively explore the polarization switching dynamics by the DP model in CIPS, it is necessary initially to confirm whether the DP model can accurately reproduce the DFT kinetic barriers of polarization switching. The out-of-plane polarization switching of CIPS originates from the movement of Cu cation normal to the layer plane. We employ the NEB method to compute the DFT kinetic barriers for the uniform polarization reversal and only one dipole flips in 40-atom unit cell where, the energy barriers are 231.2 meV/dipole and 258.1 meV/dipole, respectively. The DP model was able to reproduce these barriers accurately with 234.9 meV/dipole and 260.4 meV/dipole, as shown in Fig. 3. The similar values of two reverse modes indicates the dipole-dipole coupling is relatively weak in CIPS. Overall, the DP model accurately predicted the thermodynamic and kinetic properties of various CIPS structure, even in intermediate states.

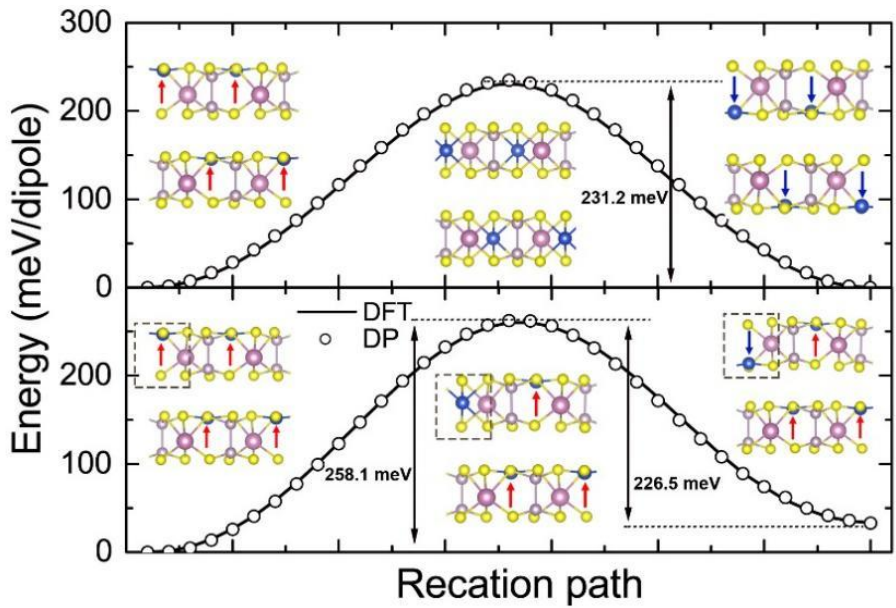

Figure 3: Nudged elastic band method calculated minimum energy path for polarization uniform reversal (a) and single Cu dipole flips (b) in bulk-CIPS by DP model and DFT. The energy barriers of above two model are very close.

## B. Temperature-driven phase transition

Using the DP model, the atomic structure and ferroelectricity of CIPS at the finite temperature can be calculated through DPMD simulations. The *NPT* MD simulations were performed to investigate the temperature-driven order-disorder ferroelectric-paraelectric phase transition, starting from the ferroelectric monoclinic phase of 38400-atom supercell. The red line in Figure 4a shows the evolution of out-of-plane lattice constant *c* with increasing temperature. Notably, there is a sudden increase of lattice constant from 13.28 Å to 13.38 Å at ~320 K, which excellently agrees with experimental results presented in figure 7(b) of the Reference[13]. Furthermore, we incorporated the quantum effect of lattice vibration in CIPS using a Quantum Thermal Bath strategy[36], similar to our recent study on quantum paraelectric strontium titanate[37]. The quantum result is marked by the blue line in Fig. 4(a). It is evident that quantum effect has a significant influence on the lattice at low temperature: the *c* axis exhibits large negative thermal expansion coefficient of $-9.85 \times 10^{-6}$ K$^{-1}$ in the temperature range below 180 K in the quantum effect calculations (see inset of Fig. 4(a)), while the *c* axis monotonously increased with the increase of temperature in the absence of quantum effect. Recently, an experiment study reported the observation of negative thermal expansion in bulk CIPS below 150 K[38]. Our DP＋Quantum Thermal Bath simulation suggests that this abnormal negative thermal expansion phenomenon can be ascribed to the nuclear quantum effect. However, it is important to note that the nuclear quantum effect only impacts the lattices below 180 K. Since the phase transition temperatures deviate far away from this range, the following calculations are based on DPMD simulations without considering nuclear quantum effect. Further comprehensive studies on the influence of nuclear quantum effect in CIPS at low temperatures will be reported in a forthcoming publication.

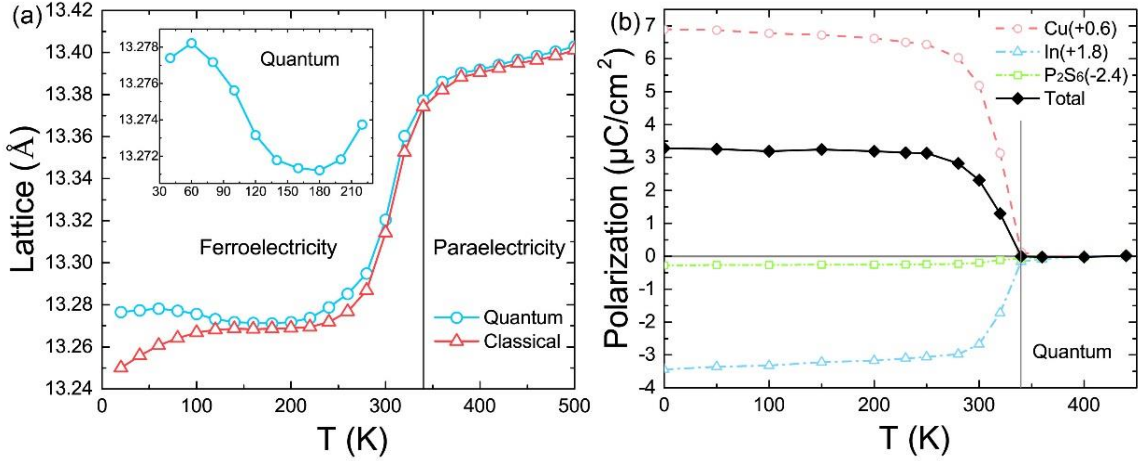

Figure 4: Out-of-plane lattice constant (a) and spontaneous polarization calculated from the DPMD simulated statistical average atomic coordinates and Born effective charge, including total and sublattice contributions (b), as a function of temperature. The inset shows a zoom-in of lattice constant at low temperature region with quantum effect. The DP model predicted $T_C$ is 340 K.

As shown in Fig. 4(a), the jump of *c* axis, both in classical and quantum cases indicates that CIPS undergoes a phase transition. To further explore the ferroelectric-paraelectric phase transition, we calculated the temperature dependence of polarization. The total polarization is determined by summing up the products of the Born effective charge and relative atomic displacement of each individual ion (see Method). The partial contribution for Cu, In, ethane-like $P_2S_6$ and total polarization as a function of temperature are illustrated in Fig. 4(b). The DP simulations predict the polarization of 3.27 $\mu C/cm^2$ below 250 K. At room temperature, the polarization is 2.41 $\mu C/cm^2$, which is in close agreement with the experimental result of 2.55 $\mu C/cm^2$. The spontaneous polarization changes rapidly in the temperature range from 300 K to 340 K, and the overall magnitude of polarization of the system decrease to nearly zero above 340 K, which clearly reveals a first-order phase transition. The DP model predicted transition temperature ($T_c$) of 340 K corresponds well with experiment value of 315 K[13].

The order-disorder characteristic of ferroelectric phase transition in CIPS is associated with intralayer Cu hopping motions. Statistical analysis of the Cu occupancies at different temperatures helps shed light on the nature of order-disorder ferroelectric phase. The DPMD simulations provide further insight into the fine structure of Cu occupancies. Figures. 5(a) and (b) display the maps of Cu occupancies in real space and the corresponding probability density contours as a function of temperature. Below 220 K, the Cu position located upper off-center site is 100% filled. Additionally, it is found that the probability distribution of Cu is asymmetry (water-drop shape) instead of a harmonic vibration around the energy minimum position[13], which is consistent with asymmetry of the electron density intensity distribution reported in a recent experiment[12]. Upon heating to 230 K, upper site of Cu cations decreases to 86 %, while the bottom site starts to be filled with 14 % Cu. Above 340 K, the probability distribution in upper and bottom sites become equivalent, yielding a centrosymmetric non-polar configuration with the appearance of a twofold axis through the octahedral center. Such dipole-disorder configuration contains Cu that randomly flip between an upper and down site within the same layer, without crossing the vdW gaps. These simulation results clearly indicate that the ferroelectric-paraelectric phase transition in CIPS is order-disorder type.

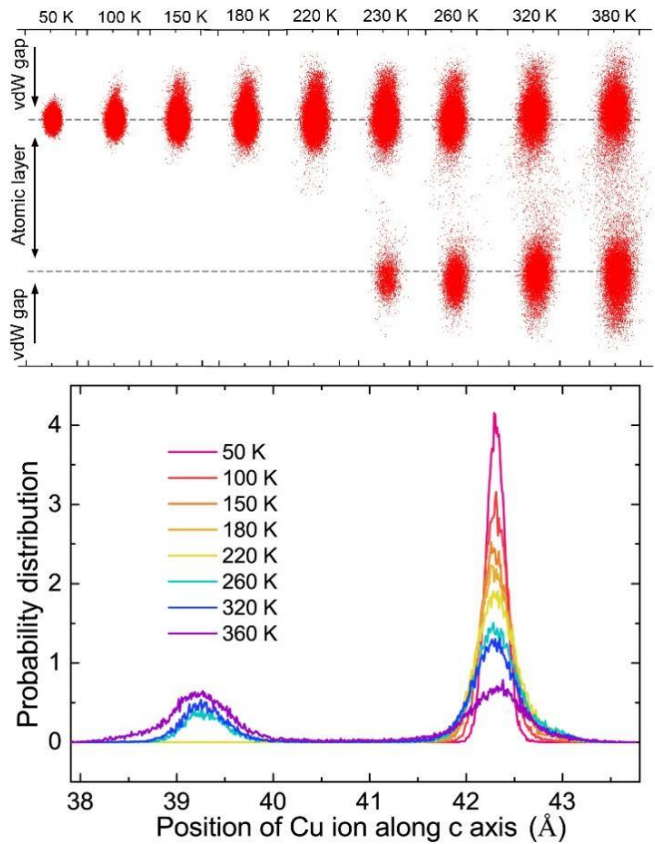

Figure 5: Temperature dependent probability distribution maps of off-center Cu cation in real space (a) and corresponding line profiles of the probability distribution along the out-of-plane (b). It indicates the distribution of Cu atoms is asymmetry (i.e. water-drop shape) instead of a harmonic vibration around the energy minimum position.

We also plot the in-plane domain pattern evolution with temperature for instantaneous configurations with 48.9 nm × 42.3 nm specimens. Figure 6 show the in-plane snapshot domain patterns in bulk system, where each pixel color represents the magnitude of the out-of-plane polarization in the corresponding unit cell. The simulation starts with single ferroelectric phase with an upward polarization. At 220 K, the single domain shows no obvious change, and the nucleation process of downward polarization domain appears at higher temperature ~260 K owing to the thermal fluctuation-driven Cu hoping motions (see Fig. 6(b)). Then the fraction of downward domains increases with temperature. At 360 K, the quantity of upward and downward domain regions is equal, forming a labyrinth-like nano-pattern with a diameter of 1 nm. These domain evolution results are consistent with the probability density of Cu dipoles in Fig. 5.

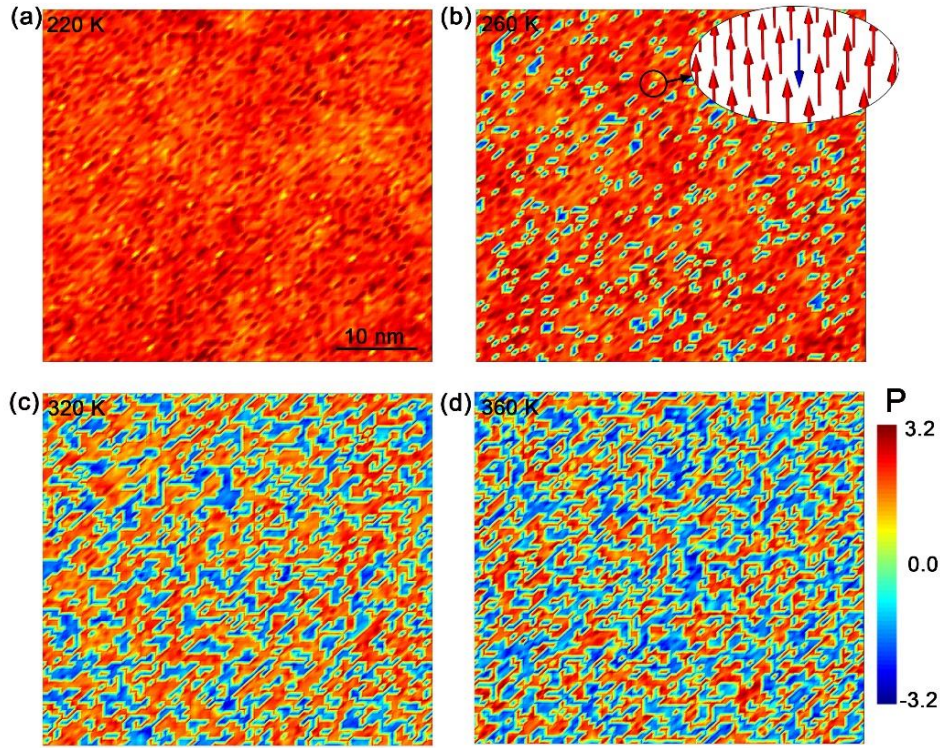

Figure 6: In-plane domain pattern for instantaneous configurations at 220 K (a), 260 K (b), 320 K (c), and 360 K (d), revealing the temperature-driven order-disorder ferroelectric phase transition in bulk CIPS. Each pixel color represents the magnitude of polarization (P) of each unit cell in the 256 000-atom periodic 80 × 40× 2 supercell.

In vdW CIPS, the ferroelectric ordering is thickness dependent. The out-of-plane ferroelectricity often vanishes below the critical thickness due to enormous depolarization field. A critical thickness of six layers (~4 nm) have been identified for the presence of ferroelectricity in CIPS flakes[4]. To further explore the ferroelectric ordering in ultrathin films, we preform DPMD simulations on bilayer and monolayer CIPS. In the bilayer system, the polarization configuration is stabilized into an interlayered head-to-head antiferroelectric state as illustrated in Fig. 7(a), although the initial configuration associated with ferroelectric monoclinic phase was used in simulations. This result agrees with the DFT calculation reslut that the head-to-head antiferroelectric ordering is the ground state for bilayer CIPS (see Table I). Such antiferroelectric ordering effectively mitigates the influence of the out-of-plane depolarization field. The antiferroelectric ordering remains highly stable in the bilayer CIPS even at temperature up to 360 K. While for monolayer system, the formation of nano-domain indicates that neither the ferroelectric nor antiferroelectric ordering can be stabilized at any temperatures as shown in Fig. 7(b). Our theoretical findings show the macroscopic ferroelectricity disappeared in both bilayer and monolayer CIPS.

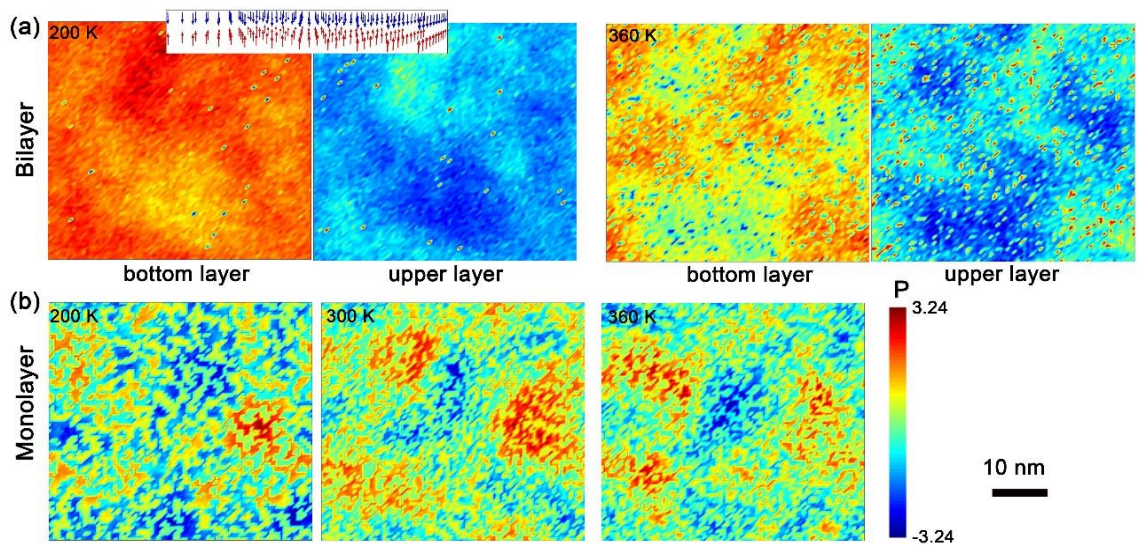

Figure 7: In-plane domain pattern characterization of bilayer (a) and monolayer (b) CIPS at different temperatures.

**C. Polarization switching dynamics**

Comprehending the ferroelectric switching dynamics in ferroelectric materials, particularly under the influence of an external electric field, is crucial for understanding the read-write speed of ferroelectric memory devices. For perovskite ferroelectrics, it is well established that the electric field-driven domain wall motion plays a critical role in ferroelectric switching process[19,20]. For example, in the bulk system of $PbTiO_3$ (PTO) preexisting 180° domain wall, the domain patterns obtained from the DPMD simulations during the ferroelectric switching process are shown in Fig. 8a. It is observed that upon the application of a small external electric field of 10 MV/m, the domain walls gradually move towards the reversal domain, and the domain walls approach closer to each other and eventually merger to produce a single domain after 50 ps. It indicates that the conventional ferroelectric switching occurs through a typical domain wall motion, whereas the behavior of CIPS differs from this mechanism. We focus on the in-plane domain evolution under an out-of-plane electric field. The applied field, slightly above the critical value, is relatively small (20 MV/m). The snapshots of the domain pattern under positive and negative electric field are shown in Fig. 8b. It is observed that the new domains nucleate within the reversed domain during the switching process at 5 ps, demonstrating that the polarization switching in CIPS is mediated by single dipole unidirectionally flip, rather than domain wall motion. As more new domain nucleus are formed, the reversed domain nearly disappears at 60 ps.

This mechanism underlying the polarization switching dynamics under an electric field in CIPS is significantly distinct from the conventional mode. The observed unconventional switching dynamics can be understood from two aspects: i) the competition between barriers of domain wall motions, single dipole flips in monodomain (i.e. nucleation), and polarization uniform reverse; ii) magnitude of domain wall energies.

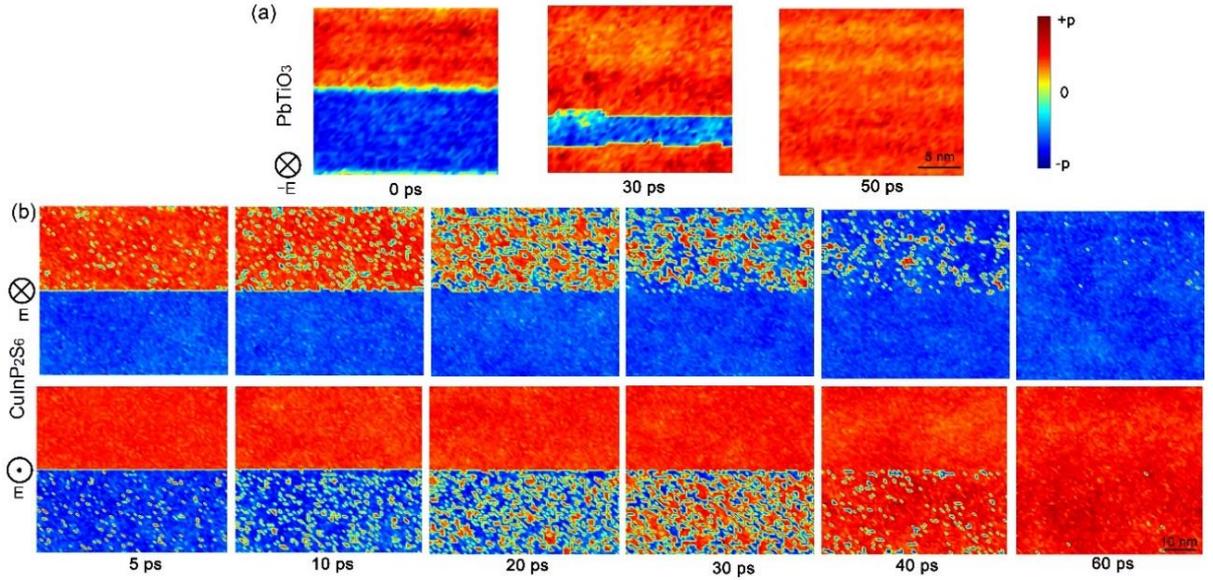

Figure 8: In-plane domain pattern revealing the ferroelectric switching of multi-domain with 180° domain wall under electric field at 220 K in PTO (a) and CIPS (b).

Firstly, we investigated the domain wall motion in PTO and CIPS between the two/three equilibrium states by performing NEB calculations for each. Figures 9(a) and (c) show the paths of domain wall motion in both PTO and CIPS, denoted by black arrows, where the 180° domain walls move from the initial state (solid line) to final state (dashed line). With this process, one of dipole (dashed box) flips. Figures 9(b) and (d) show energy profiles along the minimum energy path for the domain wall motion (blue lines), polarization uniform reverse (red lines), and single dipole flips in monodomain (gray line) for PTO and CIPS, respectively. In PTO, the energy barrier of domain wall motion is 23.5 meV, which is six times smaller than that of polarization uniform reverse (138.3 meV). In addition, the structure of single dipole flips in monodomain is unable to stabilize even at 0 K, demonstrating the nucleation of domain is energetically costly. These findings indicate that domain wall motion is the preferred mode of polarization switching in PTO due to its lower energy cost, in agreement with the domain wall motion mediated polarization switching observed in Fig. 8(a). While for CIPS, the energy barriers for domain wall motion, uniform reverse, and single dipole flips have similar magnitudes, as shown in Fig. 9(d), indicating that domain wall motion and nucleation consume almost the same amount of energy. Moreover, the structure of single dipole flips in monodomain is stable, as shown in inset of Fig. 9(d).

Therefore, the domain wall motion is not the preferred mode for polarization switching in CIPS, and nucleation will randomly occur within the entire reversed domain region.

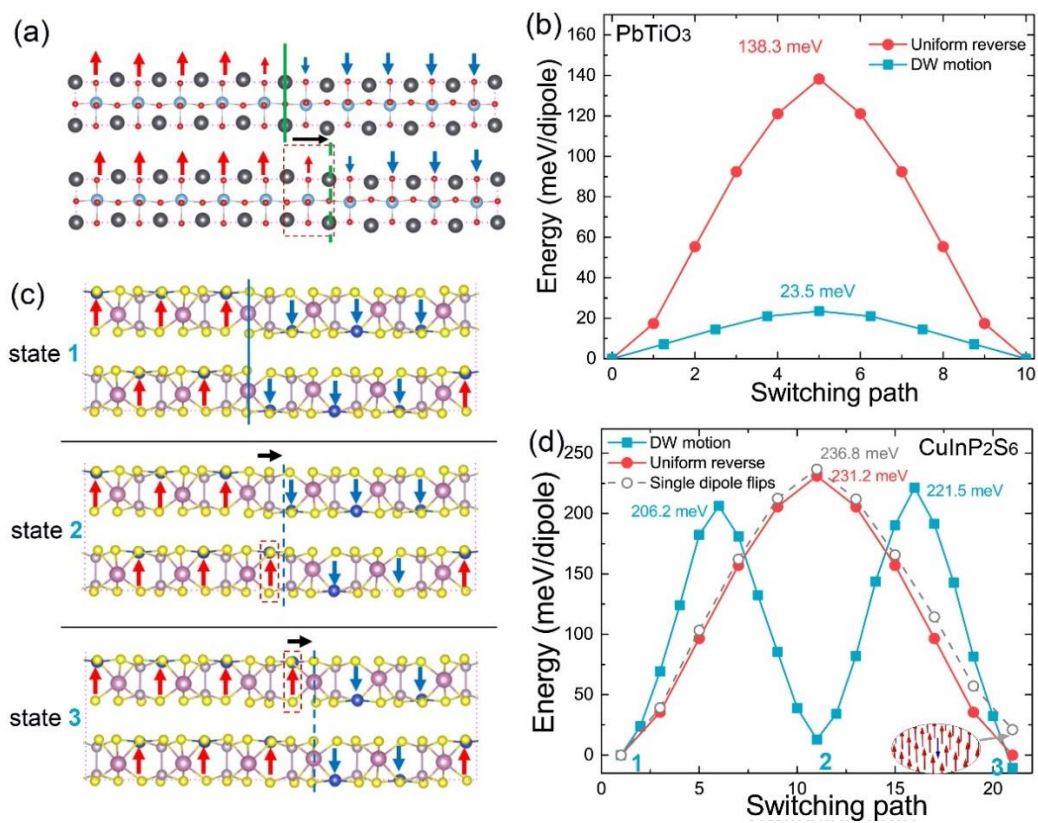

Figure 9: Energy barriers for the polarization switching governed by domain wall motion, uniform reverse, and single dipole flip. The 180° domain wall motions for PTO and CIPS are illustrated in (a) and (c), respectively. The energy profiles along the minimum energy path for the domain wall motion (red lines), polarization uniform reverse (blue lines), and single dipole flip for PTO and CIPS are shown in (b) and (d). The structure of single dipole flip in PTO is unable to stabilize.

Secondly, the magnitude of domain wall energy is also critical to polarization switching mode. The energy of 180° domain walls in PTO (see Fig. 9(a)) is 9.89 meV/Å$^2$, indicating that the formation of an abundance of domain walls during nucleation will raise the total energy of the system, resulting in the instability. The probability of domain wall motion is much higher than that of domain nucleation inside the reversed domain environment, because the former case does not increase the volume of domain wall, ensuring thermodynamic stability of the system. By contrast, the domain wall energy in CIPS (see Fig. 9(c)) is only 0.014 meV/Å$^2$, three orders of magnitude smaller than that in PTO. Such low domain wall energy ensures that the system can maintain low energy states even when an abundance of domain walls form during the flipping of many single Cu dipoles, as shown by the domain pattern of 10 ps in Fig .8(b).

## Conclusions

In conclusion, employing a machine-learning potential trained from the DFT dataset, we revealed an unconventional polarization switching mechanism in vdW ferroelectric CIPS, involving the flip of single dipole instead of the conventional domain wall motion. This unusual single dipole flip scenario is attributed to the competition between the energy barrier of domain wall motion and single dipole flips. The mechanism demonstrated in this paper provides an alternative perspective on understanding the field-driven ferroelectric switching, ionic conductivity, and other related properties in vdW ferroelectric materials. The results obtained through our DPMD simulations, such as the temperature-driven order-disorder ferroelectric-paraelectric phase transition, and the out-of-plane thermal expansion coefficient, are in agreement with experimental observations. These findings pave the way for further exploration and exploitation of vdW ferroelectrics in various applications.

All the input files, training datasets, and DP model files to reproduce the results contained in this paper are available in *AIS Square* database web[39].

**ACKNOWLEDGMENTS**

This work was supported by the National Key R&D Program of China (Grants No. 2021YFA0718900, No. 2022YFA1403000, No. 2021YFE0194200), the Key Research Program of Frontier Sciences of CAS (Grant No. ZDBS-LY-SLH008), the National Nature Science Foundation of China (Grants No. 11974365, No. 12204496, No 12161141015), the K.C. Wong Education Foundation (Grant No. GJTD-2020-11), and the Science Center of the National Science Foundation of China (Grant No. 52088101). H.W. acknowledges the support from the Zhejiang Provincial Natural Science Foundation of China (grant no. LDT23F04014F01).